# Origin of the metallic to insulating transition of an epitaxial Bi(111) film grown on Si(111)

F. Pang, X.J. Liang[a)], Z.L. Liao, S.L. Yin, and D.M. Chen[b)]

*Beijing National Laboratory for Condensed Matter Physics, Institute of Physics, Chinese Academy of Sciences, Beijing 100190, China*

**ABSTRACT**

Transport characteristics of single crystal bismuth films on Si(111)-7×7 are found to be metallic or insulating at temperature below or above $T_C$, respectively. The transition temperature $T_C$ decreases as the film thickness increases. By combining thickness dependence of the films resistivity, we find the insulating behavior results from the states inside film, while the metallic behavior originates from the interface states. We show that quantum size effect in a Bi film, such as the semimetal-to-semiconductor transition, is only observable at a temperature higher than $T_C$. In addition, the metallic interface state is shown to result from the large SO-splitting at interfaces.

Corresponding Author:

**Main text**

Bismuth (Bi) is a semimetal with extremely long Fermi wavelength of ~ 40 nm which makes it a very promising material to exhibit strong quantum size effect (QSE).[1] In early 1960s, the first observation of QSE was made in the Bi films.[2] Shortly after, Sandomirskii predicted that a semimetal-to-semiconductor (SMSC) transition would happen in Bi films when its thickness is less than approximately 20–30 nm.[3] This prediction has been investigated extensively for more than 30 years since then and the conclusion remains uncertain to date.[4] Two important factors make the observation of the SMSC transition very difficult.[5, 6] One was the difficulty in controlling the thickness of the Bi films in the early experiments, while the other has to do with the film quality as the defect-induced carriers in a poor quality film could exclude a clear signature of the SMSC transition. As a result, whether the SMSC transition can really happen has been debated up to now.

More recently, the p-type metallic surface states resulting from the large spin-orbit (SO) splitting were found on the Bi(111) surface of both bulk crystal [7] and the epitaxial film grown on Si(111)-7×7 substrate [8] by the angle-resolved photoemission spectroscopy (ARPES) measurements. Hirahara *et al.* found no Fermi-level crossings for the states inside the films, suggesting that the electrical properties of the films were governed by highly metallic surface states.[8] But there is still a lack of a direct electronic transport experiment to verify whether it really the case. The SO splitting is caused by Rashba effects due to the lack of inversion symmetry at surface/interface.[9] The strength of splitting is given by Hamiltonian $\hat{H}_{SOC} = \frac{\hbar^2}{4m^2c^2}\vec{\sigma}\cdot(\nabla V \times \vec{p})$, where $\vec{\sigma}$ is a Pauli matrix, $\vec{p}$ the momentum and $\nabla V$ the electric field which is determined by the extra electric field [10-12] and the field near the atomic nucleus[12-16]. The studies show when a surface covered by other material, the extra electric field [11,12] or the field near the atomic nucleus[12] can be changed and so can the SO splitting strength. The studies show that when a surface covered by other material, the $\nabla V$ near the atomic nucleus [11, 12] and the SO splitting strength can change. Other experimental findings, however, showed no change such as a Tungsten (W) surface covered with Au and Ag. [16] There still is no evidence on whether the $\nabla V$ is changed for a Bi(111) surface when it is covered by a non-magnetic metal.

Hirahara et al.[5] and Wells et al.[6] independently measured the conductance of Bi(111) surface states and found a large surface conductivity at room temperature. Furthermore, Behnia et al. [17] found fractional quantum Hall effect in a thick Bi film, and Seradjeh et al. [18] attribute this to the metallic surface (interface) states. Apparently the influence of the surface/interface states on the transport property of the Bi film remains an open question. In this work, we carried on a systemic investigation of the high quality epitaxial single crystal Bi(111) films covered by Al and the results clearly indicate the metallic interface states play very important roles in its electronic transport.

The single crystal Bi(111)/Si films were prepared in a homemade ultrahigh vacuum system.[19] The Bi films were grown on n-type Si(111)-7×7 structure at room temperature (RT) with a rate of 0.18 ML/min and characterized *in situ* by STM and reflection high-energy electron diffraction (RHEED). The films underwent a phase transformation from the pseudo-cubic {012} into the hexagonal Bi(111) during the growth as described in Ref. 20. To achieve flat single crystal Bi thin films, the samples were moderately annealed around 400 K. Fig. 1(a) and (b) show typical STM topographic images of the Bi film and Fig. 1(d) and Fig. 1(e) show the corresponding RHEED patterns. The observation of Kikuchi lines in RHEED pattern indicate that the entire volume of the films was long-range ordered, which was also confirmed by *ex-situ* Scanning Electron Microscope (SEM). After the growth of Bi films, ~4 ML Al was deposited on the film. Then it was tailored to a Hall-bar device by submicron ultraviolet mask and Ar+ plasma etching, followed by the lift-off process. Fig. 2(a) is a transmission electron microscopy (TEM) image of the cross section of the Hall device. It reveals that the cap Al layer and the buried interface between Bi and Si are amorphous, but the Bi film maintains good crystalline structure. Fig. 2(b) shows the SEM image of the Hall-bar device with a length ($L$) -to-width ($W$) ratio of three. The resistance measurements were performed by a physical property measurement system (Quantum Design, model 6000) in the standard four electrode method with a temperature range of 2 K to 300 K and a magnetic field up to 9 T. Fig. 2(c) shows the measured I-V curves of a 5 nm thickness film at 2 K and 300 K. The voltage varies linearly with the current, which indicates that the contacts were Ohmic. And the linear I-V curves show that the Bi films are very well isolated from the substrate by a Schottky barrier. A relatively low current (5 μA) was applied to avoid resistive heating of the film.

Generally speaking, when neglecting carriers scattering in between Bi film and Al layer or Si substrate, we can write the longitudinal conductivity $\sigma_{xx}$ as $\sigma_{xx} = \sigma_{Al} + \sigma_{Si} + \sigma_{Bi}$, where $\sigma_{Al}$, $\sigma_{Si}$ and $\sigma_{Bi}$ are the conductivity of Al layer, Si substrate, and Bi film, respectively, $\sigma_{Bi}$ consists of the contributions from Bi/Si and Bi/Al interface states, $\sigma_{interface}$, and the states inside Bi film, $\sigma_{film}$

$$\sigma_{Bi} = \sigma_{interface} + \sigma_{film} \quad (1)$$

To estimate the relative contributions from Si and Al layer, we first measured the resistance of a Si(111) substrate covered by ~4 ML Al using the same method. The resistance is about 10 kΩ at RT and its temperature dependence is semiconductor-like. Furthermore, Bi is immiscible with either Al or Si, thus the current transport in a Al/Bi/Si film is basically confined in the Bi films and $\sigma_{xx}=\sigma_{Bi}$ is a very good approximation for our experiments.

Fig. 3 shows the temperature dependence of the longitudinal resistivity, $\rho_{xx}(T)$, for films with different thickness in which the $\rho_{xx}$ is calculated from resistance $R_{xx}$ by $\rho_{xx} = R_{xx}\left(W/L\right)$. Three interesting characters can clearly be found: (1) a transition temperature, $T_C$, exists in each film. Above $T_C$ the resistivity increases as temperature decreases and below $T_C$ the relation is reversed, showing metallic and insulating behaviors below or above $T_C$; (2) Thinner film yields higher $T_C$ and a larger slope of $\rho_{xx}$ vs. $T$ on the metallic side, or a stronger metallic character; and (3) For the ultra thin film (≤10 nm), their resistivity converges at very low temperatures.

Based on Drude model, the conductivity of a film can be expressed as $\sigma = \sum n_i e \mu_i$, where i represents electron or hole, $n_i$ the carrier density, and $\mu_i$ the mobility. For the two dimensional case $n_i = n_{2D}/d$, $n_{2D}$ is 2-dimensional carrier density, $d$, the thickness of the film. It is well known that when the thickness of a film becomes thin, $\sigma_{film}$ will decrease owing to the reduction in carrier density and mobility resulting from the enhancement of the interface scattering. Here we assume that $\sigma_{interface}$ does not depend on the film thickness under the same consideration as in Ref. 5. As a result, when the film becomes thin the contribution of $\sigma_{interface}$ becomes dominant in $\sigma_{Bi}$ due to the decreasing contribution from $\sigma_{film}$. Consequently, the rise of $T_C$, the enhancement of metallic character, should be attributed to the influence of the interface states. The convergence of the resistivity at low temperature in Fig. 3 further suggests that the interface states are metallic. The resistance in low temperature regimes can be fitted approximately by

$$\rho_{interface} = \rho_0 + \alpha T \qquad (2)$$

where $\rho_0$ and $\alpha$ are the fitting constants. Eq. (2) is a typical resistivity expression for a metal when $T \gg \Theta_D$, $\Theta_D$ being the Debye temperature. A recent study showed that $\Theta_D \sim 33$ K [21] for the surface state of Bi film on Si substrate which is much lower than in bulk Bi[1] due to the decrease of the surface atom binding energy. For the thinnest film (5 nm) in Fig. 3, $\sigma_{Bi}$ should be dominated by $\sigma_{interface}$ and increases linearly with $T$ as is found in Fig. 4(a). Therefore, the metallic behavior of the thicker films in Fig. 3 should also result from the interface states.

As for the insulating behavior in the high temperature regime it should result from the states inside the film. In a very thin single crystal film, the mean free path of the carrier inside film mainly depends on the film thickness, so the mobility is nearly independent of temperature.[22] Consequently the temperature dependence of conductivity reflects the temperature dependence of the carrier concentration. From the result of Hirahara's et al.[8], it is reasonable to conclude that the electronic states inside films lie below Fermi level and so the mobile carriers mainly result from thermally excitation. This leads to a conductivity expression as

$$\sigma_{film} = \sigma_0 \exp\left(-\frac{E_g}{2kT}\right) \qquad (3)$$

where $\sigma_0$ is a constant, $E_g$ the energy band gap and $k$ the Boltzmann constant. In addition, the decreasing of $T_C$, or a relatively larger insulating segment and a stronger exponential dependence for the thicker films in Fig. 3, also support that the insulating behavior results from the states inside the film.

To summarize, the temperature dependent resistivity can be divided into insulating and metallic regime separated by $T_C$. In the insulating regime ($T > T_C$), the conductance in Bi film layer $\sigma_{film}$ dominates the transport, while in the metallic regime, the conductance in the Bi film layer becomes quite poor and conductance via the interface states dominates the transport. We thus combine equations (2) and (3), $\rho_{xx} = 1/(\sigma_{interface} + \sigma_{film})$, to fit the temperature dependent resistivity curve measured for a 8nm Bi film and the agreement is quite reasonable as shown in Fig. 4(b).

The above discussions show that $T_C$ is a good physical parameter which characterizes the transport behavior of Bi film. Below $T_C$, the Bi film behaves as a metal while above $T_C$ it is an insulator. Thus for a Bi film with a thickness lower than 30nm, the insulating phase induced by QSE can be observed only at a temperature above $T_C$. Fig. 4(c) plots the $T_C$ as a function of the Bi film thickness. We note that for $d = 5$ nm, $T_C$ is very close to room temperature. That is, a Bi film (<5 nm thick) is metallic even at RT. Therefore, there exists only a very narrow window to observe the SMSC transition for Bi films thinner than 30 nm at a temperature below *RT*.

Apparently the QSE inside film will induce the decreasing of the density of states around Fermi level. Indeed T. Hirahara *et al.* [8] found the electron state density inside Bi films below $E_F$ sharply decreased with the thickness due to QSE. The enhancement of insulating behavior with the thickness decreasing may be result from the influence of QSE on the transport. Generally the energy levels determined by QSE can be written as $-\Delta (n/d)^2$, where n>0 is integer, $\Delta=\hbar^2\pi^2/2m^*$ ($\hbar$, $m^*$ plank constant and the carrier effective mass), $d$ the thickness of films.. So if QSE induces the insulating behavior, the $E_g$ should be scaled by $1/d^2$. And then we can obtain the thickness dependence of $T_C$ by $\frac{\partial \rho_{xx}}{\partial T}=0$, as indicated in Fig. 4(c) (red line). Consequently, QSE is one of possible cause for the insulating states inside films.

By extrapolating the data for the 5 nm Bi film, we can get an estimation of lower bound of the interface state resistivity, $\rho_{interface} = 548\Omega/\square$ at RT shown in Fig. 4(a) as indicated by red broken line. The true resistivity for the interface states should be a little larger to account for the influence of $\sigma_{film}$. The interface state resistivity obtained by our measurements is basically the same as the surface state resistivity reported in Ref. 5. This suggests that the SO splitting has little difference between Al/Bi and vacuum/Bi though the potential gradient at the Al/Bi is much smaller than the vacuum/Bi based on the consideration of work function. Consequently, SO coupling, or, $\nabla V$ should mainly depend on the field near the atomic nucleus. This is also consistent with the atomic structure of the Bi film. The single crystal Bi(111) film is terminated with a bilayer structure with saturated covalent-like bonds. As shown in Fig. 2(a), the amorphous structure of the Ai/Bi interface should results in a low Al-Bi coordination and hence less likely to cause a surface core level shift as in the case of Li adsorbed on W.[13] As a result, the Al atoms adsorbed on Bi film does not react with Bi atoms and cannot induce a surface core level shift as in the case of Li adsorbed on W.[12] The Al/Bi structure is more similar to that of Au/W or Ag/W [16] where the SO splitting is independent of the coverage of Au or Ag.

Furthermore, the magnetic field dependence of the resistance for different thickness film shows that there is still a very strong SO coupling at the interface and the carriers from interface states are p-type, which is the same as the conclusion from the ARPES results obtained on vacuum/Bi/Si.[8] Fig. 5 shows the magnetic field dependence of the Hall resistance of our device and it indicates that the films are p-type and as the film thickness increases the carrier type approaches to a change from p-type to n-type as one would anticipate since bulk Bi is n-type.[1, 23] Therefore, we conclude that the metallic interface states result from strong SO splitting at the interface, which is not influenced by Al layer basically.

In conclusion, we have shown that the carrier transport in a thin single crystal Bi(111) film can be characterized by a transition temperature $T_C$ which strongly depends on the film thickness. Below $T_C$ the interface states dominate the conductance and the film behaves metallic. Above $T_C$ on the other hand, the states inside film dominate the conductance and the film is insulating. Thus QSE in a Bi film, such as the SMSC transition, is only observable at a temperature higher than $T_C$. In addition, the metallic interface state is shown to result from the large SO-splitting at interfaces, not destroyed by Al layer coverage.

**Acknowledgement**

This work was supported by National Natural Science Foundation of China under grant No. 10874217 and No. 10427402 and a grant from the National Basic Research Program of China (973 Program) No. 2006CB933000. We thank Prof. Z. Fang, Y. Q. Li and Dr. Z. Z. Wang for their invaluable discussions and Dr. H. F. Yang for the help in microfabrication.

**Figure Captions:**

**Fig. 1**. (color online) STM images (150 nm × 150 nm, -2.0 V) of 4 ML Bi{012} film (a) and 14 ML Bi(111) film (b), (c) and (d) the corresponding RHEED patterns.

**Fig. 2**. (a) the cross section TEM image of the fabricated device; (b) SEM image of a Hall bar device; (c) four-electrode *I-V* curves taken at 2 K and room temperature respectively.

**Fig. 3**. (color online) Temperature dependent resistivity of the Bi films with different thickness.

**Fig. 4.** (color online)  (a) The temperature dependent resistivity for the 5 nm (a) and 8 nm (b) thick Bi films (black circle). The red solid line is the result of fitting; (c) The transition temperature $T_C$ as a function of the thickness (black circle) and the red line is a guide to the eye as described in the text.

**Fig. 5.** Thickness dependence of the low-field Hall coefficients at 2 K.

**Fig. 1**

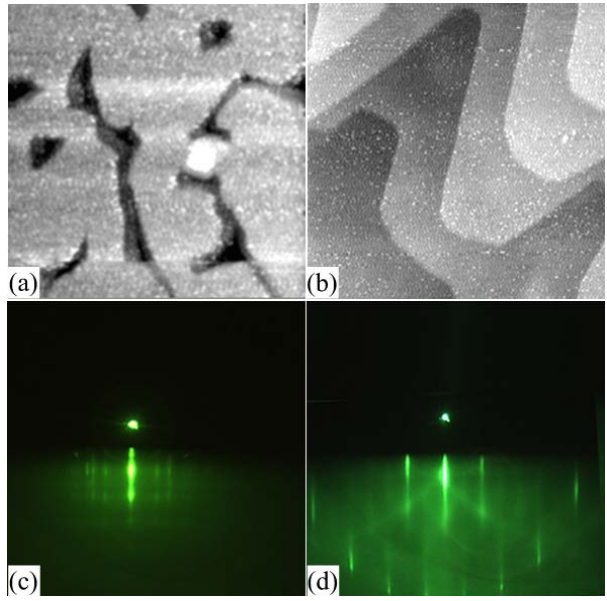

**Fig. 2**

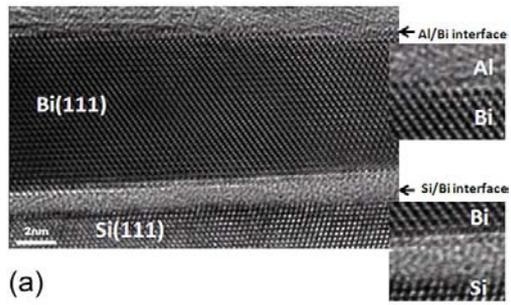

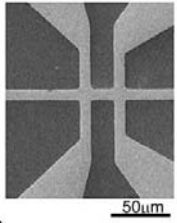
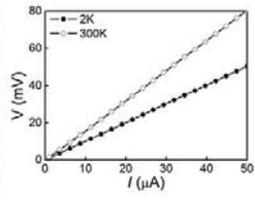

(a)

(b)  (c)

**Fig.3**

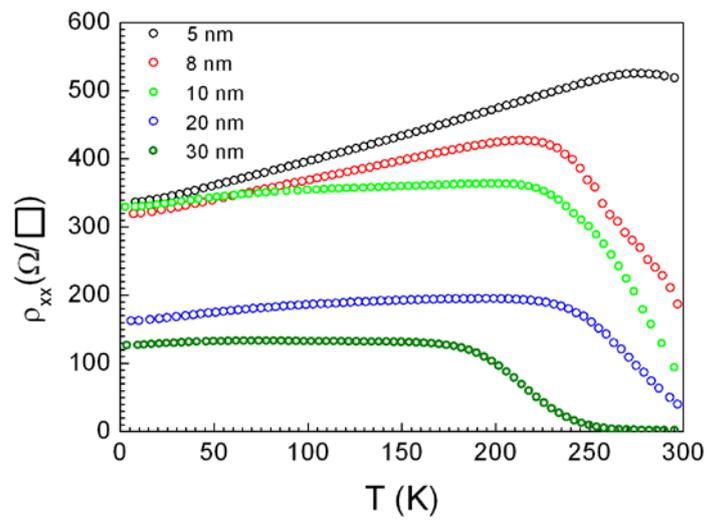

**Fig.4.**

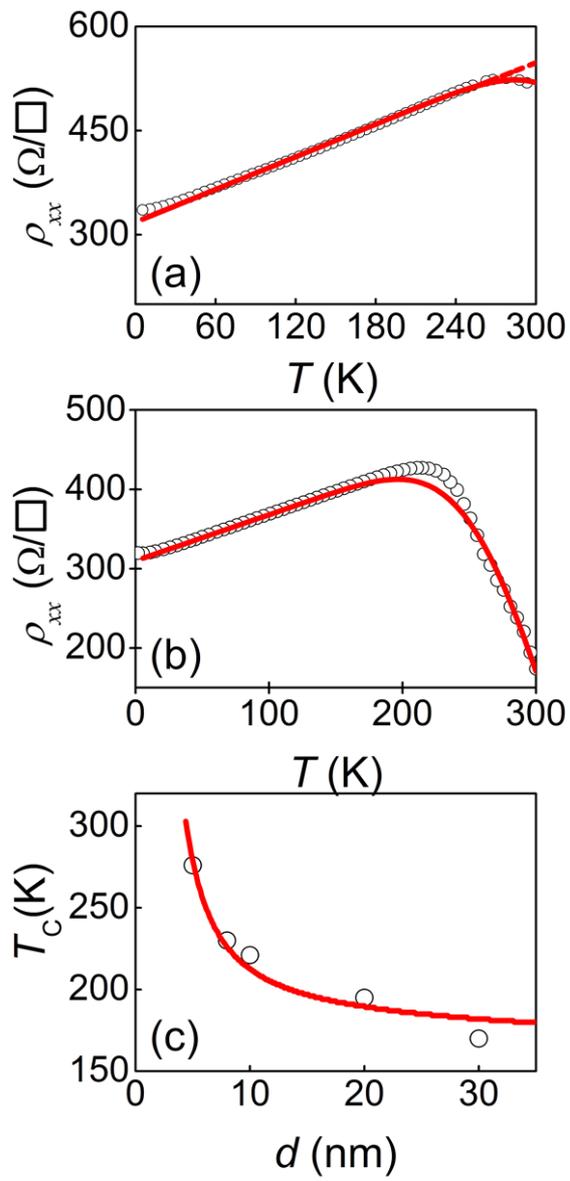

**Fig.5.**

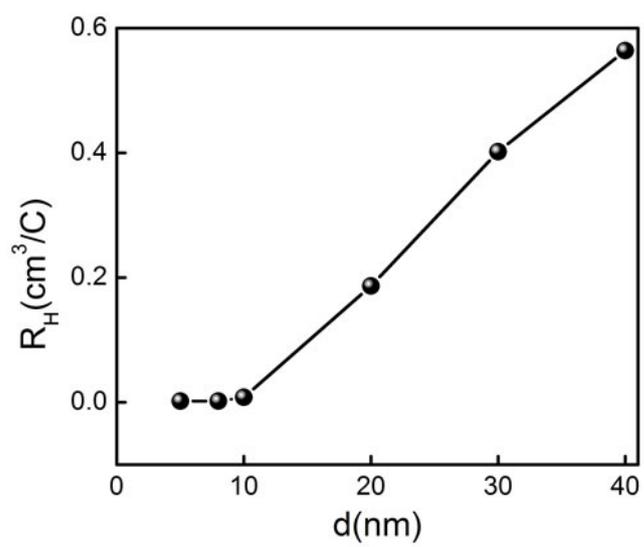